\documentclass[aps,pra]{revtex4}
\usepackage{epsfig}
\usepackage{amsfonts,psfrag}
\newcommand{\be}{\begin{equation}}
\newcommand{\bea}{\begin{eqnarray}}
\newcommand{\ee}{\end{equation}}
\newcommand{\eea}{\end{eqnarray}}
\newcommand{\eq}[1]{Eq.~\ref{#1}}
\newcommand{\eqs}[2]{Eqs.~\ref{#1} and \ref{#2}}

\newcommand{\ud}{{\rm d}}
\newcommand{\ue}{{\rm e}}

\begin{document}
\title{Casimir force between integrable and chaotic pistons}
\author{Ezequiel \'Alvarez$^1$, Francisco D.\ Mazzitelli$^1$, Alejandro G.\ Monastra$^2$,  and Diego A.\ Wisniacki$^1$}

\affiliation{ $^1$ Departamento de F\'\i sica, FCEyN UBA, and IFIBA, CONICET,
Ciudad Universitaria, Pabell\' on I, 1428 Buenos Aires, Argentina}

\affiliation{$^2$ Gerencia de Investigaci\'on y Aplicaciones, Comisi\'on Nacional de Energ\'\i a At\'omica, CONICET, Av.\ Gral.\ Paz 1499, 1650 San Mart\'{\i}n, Argentina}

\begin{abstract}
We have computed numerically the Casimir force between two identical pistons inside a very long cylinder, considering different shapes for the pistons.
The pistons can be considered as quantum billiards, whose spectrum determines the vacuum force.
The smooth part of the spectrum fixes the force at short distances, and depends only on
geometric quantities like the area or perimeter of the piston. However, correcting terms to the force, coming from the oscillating part of the spectrum which is related to the
classical dynamics of the billiard,  are qualitatively different for classically integrable or
chaotic systems.
We have performed a detailed numerical analysis of the corresponding Casimir force for pistons
with regular and chaotic classical dynamics.
For a family of stadium billiards, we have found  that the correcting part of the Casimir force
presents a sudden change in the transition from regular to chaotic geometries.
\end{abstract}

%\pacs{PACS number(s) go here}

\maketitle

%%%%%%%%%%%%%%%%%%%%%%%%%%%%%%%%%%%%%%%%%%%%%%%%%%%%%%%%%%%%%%
%%%%%%%%%%%%%%%%%%%%%%%%%%%%%%%%%%%%%%%%%%%%%%%%%%%%%%%%%%%%%%
\section{Introduction}

In the last years, there has been an increasing interest in the Casimir effect \cite{reviews}. The Casimir force,
which in its simplest version consists in the interaction between two perfectly conducting surfaces, involves the
summation of the vacuum energy of the electromagnetic modes allowed by those surfaces, and therefore depends
in a complex way on the spectrum associated to the considered geometry.

In this paper we will consider the following system: a pair of identical pistons of arbitrary shape separated by a distance $a$,
contained in a very long cylinder with the same transversal shape \cite{pistons}. All surfaces will be considered to be perfect conductors. This system has been solved, in the sense that there is an exact formula that relates the (renormalized) Casimir energy or force with the spectrum of the two-dimensional Laplacian operator constrained to the
geometries that define the pistons \cite{marachevsky}.
The exact formula, first derived in \cite{marachevsky} (see \eq{mara} below),
involves a summation of functions of the eigenvalues of the two-dimensional
Laplacian. This summation is different from the usual sum over modes that defines formally the vacuum
energy, since the latter involves the eigenvalues of the three dimensional problem and gives a convergent result
only after a proper subtraction. The (formal) sum over modes is not adequate to implement direct numerical evaluations of the Casimir energy  even in the simplest situations \cite{Rodrigues}, because of its highly oscillatory behavior. However, as we will see in many specific examples, this is not the case with the exact formula for pistons.

With the possibility to compute the Casimir force using the spectra of two-dimensional pistons, considered as
quantum billiards, let us ask a natural question: is there any signature of the classical dynamics on the billiard in the Casimir effect? This conceptual question is based on the fact that spectra of chaotic and non-chaotic billiards are qualitatively different. Whereas the spectrum of a classically chaotic system shows level repulsion and its statistical properties are well described by Random Matrix Theory (RMT) \cite{BohGiaSmi1984}, the energy levels of a regular classical system are Poisson distributed \cite{BerTab1977} (see \cite{qchaos} for a general discussion).
Moreover, the spectrum of a given geometry can be written as the sum of a {\it smooth} plus an {\it oscillating} part. The smooth contribution to the spectrum depends only on
geometric quantities like the area, perimeter, and the curvature of the border. But the oscillating part depends on quantities related to the classical periodic orbits of the system (as action, period and stability), which are completely different in the regular or chaotic case. Moreover, periodic orbits in regular systems come in continuous families, whereas in chaotic systems they are isolated. As already shown in Ref.\cite{Jaffe,Fulling}, the smooth part of the spectrum gives the short distance behavior of the force between pistons, the leading term being the well known Proximity Force Approximation (PFA). Therefore, any signal of quantum chaos, if present, should pop out in the difference between
the full Casimir force and its smooth part.

The goal of this paper is to analyze the Casimir force between pistons trying to see some
signature of chaos. For this reason we have computed the force between pistons of different shapes
with regular and chaotic classical dynamics. In order to compute the Casimir force in our systems, it is necessary to handle summations of functions of the eigenvalues of the considered geometry.
For very simple geometries, like a rectangle or equilateral triangle, the eigenvalues have explicit analytic expressions in terms of two quantum numbers, and the computation is relatively easy (although, as we will see, a large number of eigenvalues are necessary in order to reproduce the PFA limit).

For circles, the eigenvalues are defined as the zeros of Bessel functions. In this case, one can follow two different approaches: to compute numerically the eigenvalues, and then perform the summation, or, alternatively, to perform the sum over the eigenvalues by using the argument theorem.

Besides, for  rectangular and equilateral triangular billiards, the correcting part of the Casimir force, that comes from the oscillating part of the spectrum, can be computed using exacts results from semiclassical periodic orbit theory \cite{triangle}. We remark that semiclassical techniques have been already used in the computation of Casimir force \cite{MazSanScoSte2003}. For more complex geometries, semiclassical theory is only approximated and the effort to compute the periodic orbits is enormous. The only alternative is to find the eigenvalues associated to a given geometry using sophisticated methods. We will show that, with some computational effort, the exact formula is useful to evaluate the Casimir interaction between pistons for both integrable and non-integrable geometries.

The paper is organized as follows. In Section II we review the main results for the force between pistons. We include a discussion of the exact formula, the behavior of the Casimir energy at large distance, and the relation between the short distance behavior and the smooth part of the spectrum.
In Section III we compute the Casimir force between pistons with integrable classical dynamics. We will consider rectangular, triangular and circular pistons. We will use the large and short distance behaviors as benchmarks for the numerical calculations, and show that, up to a given accuracy, it is possible to find analytic expressions for the force for any distance between the pistons.
We will also compute the correcting part of the force, and find that circular pistons have a different behavior in the short distance limit: the leading contribution to the correcting part of the force is finite for triangles and rectangles, and diverges for circles.
In Section IV we present the results for the force between pistons that are billiards with chaotic classical dynamics. The calculation of the Casimir force in this case was done by computing the corresponding eigenvalues through the {\it scaling method} \cite{vergini}.  In particular, we will analyze a family of stadium billiards,  and the uniformly hyperbolic Sinai-type billiard introduced in Ref.\cite{weird}.
 For each considered geometry, we will compute the full
Casimir force, paying particular attention to its correcting part. We
will show that at short distances all geometries with curved boundaries present a similar divergent behavior.
Section V contains an analysis of the Casimir force associated to the oscillating part of the spectrum for both integrable and chaotic pistons. The analysis is based on the computation of the correcting part of the force for a family of billiards that interpolates between a regular geometry (quarter of circle) and chaotic billiards. We will see that, in the short distance limit, the correcting part of the force has an abrupt change in the transition from  the quarter of circle to a stadium billiard.
We present our conclusions in Section VI.

%%%%%%%%%%%%%%%%%%%%%%%%%%%%%%%%%%%%%%%%%%%%%%%%%%%%%%%%%%%%%%
%%%%%%%%%%%%%%%%%%%%%%%%%%%%%%%%%%%%%%%%%%%%%%%%%%%%%%%%%%%%%%
\section{Casimir force between pistons: the exact formula}

We consider a very long electromagnetic cylindrical cavity, with an arbitrary section (which we will assume to be simply connected).
The cavity contains two plates (pistons) separated by a distance $a$. All surfaces are perfectly conducting. The ${\bf z}$-direction is the axis of the cavity, and we will denote by ${\bf x}_\perp$ the coordinates in the transverse sections.
At classical level, the electromagnetic field admits a description in terms of independent Transverse Electric (TE) and Transverse Magnetic (TM) modes, which are defined with respect to ${\bf z}$-direction. This is
possible due to the particular geometry we are considering, that have an invariant section
along the ${\bf  z}$ axis.
As the section is simply connected, there are no Transverse Electromagnetic Modes (TEM). The relevance of TEM modes in the Casimir force between pistons
with non-simply connected sections has been analyzed in Ref.\cite{Alvarez09}.

As discussed in Ref.\cite{Alvarez09}, TE and TM Casimir energies correspond to that of a
set of massive scalar fields in $1+1$ dimensions, with masses given by the eigenvalues associated to the two-dimensional pistons,
that we will denote by $\lambda_{k {\rm N}}$ and $\lambda_{k {\rm D}}$.  More explicitly,
the eigenfrequencies associated to the TE and TM modes are
\begin{equation}
\begin{array}{rcl}
w_{k,n}^{\rm TE}&=&\sqrt{(\frac{n\pi}{a})^2+\lambda_{kN}^2}\,\,\,\,\,  (n=1,2,3,..)\nonumber\\
w_{k,n}^{\rm TM}&=&\sqrt{(\frac{n\pi}{a})^2+\lambda_{kD}^2}\,\,\,\,\,  (n=0,1,2,...),\,  ,
\end{array}
\label{freq}
\end{equation}
where the eigenvalues are defined by
\begin{equation}
\nabla^2_\perp \varphi_{\rm TE,TM} = - \lambda_{\rm N,D}^2 \varphi_{\rm TE,TM}\, .
\label{tlap}
\end{equation}
The eigenfunctions $\varphi_{\rm TE,TM}$ satisfy Neumann and Dirichlet boundary conditions, respectively, on the border of the transversal section.
The  $\varphi_{\rm TE}=const $ eigenfunction with $\lambda_{\rm N}=0$ should be excluded
because it does not  correspond to a physical electromagnetic solution.

The Casimir energy $E_m$ and force $F_m$ for a field of mass $m$ in $1+1$ dimensions has been computed previously by many authors \cite{Emassive}. They are given by
\begin{equation}
E_m(a)=-\frac{1}{2\pi} \sum_{l=1}^{+\infty}
\frac{m K_1(2l m a)}{l} \, ,
\label{emassive1}
\end{equation}
and
\begin{equation}
F_m(a)=-\frac{\partial E_m}{\partial a}=\frac{1}{\pi}\sum_{l=1}^{+\infty}  m^2 K'_1(2l m a)\, ,
\label{fmassive1}
\end{equation}
where $K_1$ is the modified Bessel function of the second kind and the prime denotes derivative with respect
to the argument.
Using these results and the analogy between the TE and TM eigenfrequencies with the eigenfrequencies of massive scalar fields in $1+1$ dimensions, we can easily obtain the TE and TM contributions to the Casimir force between
pistons in the cylindrical cavity
\be
F^{\rm TE}(a)+F^{\rm TM}(a)=\frac{1}{\pi} \sum_{l=1}^{+\infty}
\left(
\sum_{\lambda_{k {\rm N}}} \lambda_{k {\rm N}}^2 K'_1(2l\lambda_{k {\rm N}} a) +
\sum_{\lambda_{k {\rm D}}} \lambda_{k {\rm D}}^2 K'_1(2l\lambda_{k {\rm D}} a)
\right)\, .
\label{mara}
\ee
This equation has been previously obtained in Ref. \cite{marachevsky} using a different method.
We stress that the formula is valid for a cavity of arbitrary section.
In the rest of the paper
we will be mainly concerned with the particular case of Dirichlet boundary conditions (which correspond to TM modes), because this will be enough for our purposes. Therefore we will omit the subindex D in the eigenvalues and the supraindex TM in the forces.

In the next sections we will present numerical evaluations of the Casimir force between pistons using
\eq{mara}. In the cases where the eigenvalues are known analytically or numerically,  the evaluation will be performed by computing explicitly the summations over $l$ and $\lambda_k$ in \eq{mara}. There are some cases where the eigenvalues associated to a given geometry are known implicitly through the zeros of a function $f(\lambda)=0$. In this case, the sum over eigenvalues can be performed using the argument theorem
\be
\sum_{\lambda_{k}} \lambda_{k}^2 K'_1(2l\lambda_{k} a)= 2\pi i\oint_C dz\, z^2K'_1(2zla)\frac{f'(z)}{f(z)}\, ,
\label{cauchy}
\ee
where the integral is performed in the complex plane $z$ along a closed contour
$C$ that encloses all zeros
of $f(z)$.

We will now
describe some general properties of the force between pistons given in \eq{mara}. The function
$$K'_1(x)=-\frac{1}{2}(K_0(x)+K_1(x))$$
is a negative and increasing function for real and positive arguments, and vanishes exponentially
for $x\to\infty$. Therefore, the force between pistons is always attractive and decreases with distance.
Moreover, at large distances, the sum is dominated by the term with the lowest eigenvalue $\lambda_1$
and $l=1$, that is

\be
F(a) \approx F_{\infty} (a) = -\frac{1}{2\pi}\lambda_1^2\left(K_0(2\lambda_1 a)+K_1(2\lambda_1 a)\right ) \approx -\frac{1}{2}\left(\frac{\lambda_1^3}{\pi a}\right)^{1/2}e^{-2\lambda_1 a}\, .
\label{largedistance}
\ee
The exponential behavior is due to the finite size of the pistons, that produce a gap in the spectrum. The above approximation is valid as long as $a \lambda_1 \gg 1$.

The behavior at short distances requires a deeper analysis. As the distance between pistons decreases, more and more eigenvalues give a sizable contribution to the force. Taking into account the exponential behavior of the Bessel functions, if in the numerical computation one neglects terms that are smaller than $e^{-D}$ for a given accuracy, all terms with $2l\lambda_k a_{min}<D$ should be kept in the calculation, where $a_{min}$ is the minimum distance for which the force is numerically computed.

It is useful to rewrite \eq{mara} in an integral form

\be
F(a) = \frac{1}{\pi} \sum_{l=1}^{\infty} \int {\text d} \epsilon \ \epsilon \ K'_1 ( 2 l a \sqrt{\epsilon} )
\rho (\epsilon)  \ , \label{integralforce}
\ee
where $\rho (\epsilon) = \sum_k \delta (\epsilon - \lambda_{k}^2 ) = \frac{\ud N}{\ud \epsilon}$ is the density of states (or energy levels). The number of energy levels $N$ below a given energy $\epsilon$ can be approximated using the so called Weyl's theorem \cite{qchaos}, which for two dimensional billiards with Dirichlet boundary conditions gives

\be
N(\epsilon) = \left(\frac{A}{4\pi} \epsilon - \frac{P}{4\pi} \epsilon^{1/2} + \chi  \right) \Theta(\epsilon) + \widetilde{N} (\epsilon) \, , \label{WeylN}
\ee
where $A$ is the area of the piston, $P$ its perimeter, and $\chi$
is related to the shape of the boundary through
\begin{equation}
 \chi=\frac{1}{24}\sum_i\left(\frac{\pi}{\alpha_i}-\frac{\alpha_i}{\pi}\right)+\frac{1}{12\pi}\sum_j\int_{\gamma_j}
\kappa(\gamma_j)d\gamma_j\,.
\end{equation}
Here $\alpha_i$ is the interior angle of each corner and $\kappa(\gamma_j)$ is the curvature of each smooth section $\gamma_j$
of the border.
The term $\widetilde{N} (\epsilon)$ contains lower order contributions that vanish in the limit $\epsilon\to\infty$. It oscillates rapidly with energy, and  is related with the classical periodic orbits of the billiard. (For TE modes the sign of the perimeter term changes
and the factor $\chi$ should be replaced by $\chi -1$ in \eq{WeylN}).
%and the term proportional to $\chi$ can be different.

Introducing the first three terms of the Weyl expansion into \eq{integralforce}, the integral and the infinite sum over $l$ can be solved, giving

\begin{equation}
F(a) \approx F_{W} (a) = -\frac{3 \zeta (4) }{ 8\pi^2a^4 } A + \frac{\zeta(3)}{32\pi a^3} P - \frac{\zeta(2) \chi} {4\pi a^2} \ . \label{weyl}
\end{equation}
This result has already been shown, using different methods, in Refs. \cite{marachevsky,Jaffe,Fulling}. The first term  is the usual PFA proportional to the area. There are also corrections of lower order on $a$ related to the perimeter and curvature of the boundary of the  billiard, although the regular or chaotic classical dynamics on the surface does not enter in this expression.

%%%%%%%%%%%%%%%%%%%%%%%%%%%%%%%%%%%%%%%%%%%%%%%%%%%%%%%%%%%%%%%%%%%
\section{Pistons with regular classical dynamics} \label{section-regular}

In this Section we use the above results in order to compute numerically the Casimir force between pistons with rectangular, triangular and circular shapes. Thinking these shapes as billiards, they all have regular dynamics as they have two constants of motion (as required for two dimensional systems). We verify that the results converge to the expected $F_{W}$ expression \eq{weyl} for short distances and to the $F_{\infty}$ expression, \eq{largedistance}, for large distances.

%Although most of these pistons may be computed using Neumann bounday conditions, we focus in Dirichlet boundary conditions it is the one %for which exists a method \cite{vergini} to compute the eigenvalues for surfaces of any shape, which is exploited in next Section.  We %end this Section with some brief comments on the calculation using Neumann boundary conditions.

In order to compute the Casimir force from \eq{mara}, it is necessary to obtain the eigenvalues of the Laplace
equation for each geometry. For a rectangular piston of sides $L_x$ and $L_y$, these eigenvalues are trivially

\be
\lambda_{n m} = \sqrt{\left( \frac{\pi n}{L_x} \right)^2  + \left( \frac{\pi m}{L_y} \right)^2}
\label{eigen_rect}
\ee
with indices $n,m \geq 1$ for Dirichlet boundary conditions (TM modes). For Neumann boundary conditions (TE modes) one of the indices can be zero (but not both at the same time).

For an equilateral triangle of side $L$

\be
\lambda_{n m} = \frac{4 \pi }{3 L} \sqrt{ n^2 + m^2 - n \, m}
\label{eigen_tri}
\ee
with $n \geq 1$ and $m \geq n+1$ for Dirichlet boundary conditions, and $n \geq 0$ and $m \geq n$ for Neumann boundary conditions, excluding the term $(n,m)=(0,0)$, see \cite{triangle, BraBha1997} for further details about the equilateral triangular billiard.

On the other hand, for a circular piston of radius $R$, the eigenvalues of the Laplace equation are given implicitly as the solutions of
\bea
\mbox{Dirichlet b.c.} &\rightarrow& J_n (R \, \lambda_{nm}) = 0, \label{circle_D} \\
\mbox{Neumann b.c.} &\rightarrow& J'_n (R \, \lambda_{nm}) = 0, \label{circle_N}
\eea
where $n \geq 0$ is an integer which labels the order of the Bessel function, and $m$ refers to its $m$-th root. For $n=0$, the eigenvalues are non degenerated, and for $n \geq 1$ they are double degenerated. For Neumann boundary conditions the first trivial zero of $J'_0$, that is $\lambda_{01} = 0$, is excluded.
Therefore, the Casimir force can be computed using Cauchy's theorem as in \eq{cauchy}. Alternatively, the eigenvalues can be computed numerically from \eq{circle_D}, and then use \eq{mara} to calculate the force. We have used both methods (see appendix \ref{b} for details on the samples of used eigenvalues).

%%%%%%%%%%%%%%%%%%%%%%%%%%%%%%%%%%%%%%%%%%%%%%%%%%%%%%%%%%%%%%%%%%%
\subsection{Casimir force}

\begin{figure}[h]
\psfrag{x}{ $a$}
\psfrag{y}{}
\includegraphics[width=.5\textwidth]{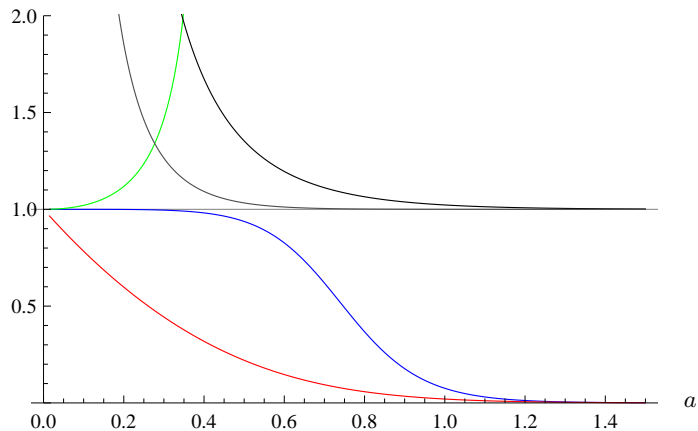}
\caption{(Color online) Ratio of the numeric Casimir force between squared-shape pistons to the Weyl expression (\eq{weyl}), $F_{num}/F_{W}$, and to the large distance asymptotic expression (\eq{largedistance}), $F_{num}/F_{\infty}$.  For those converging to $1$ at small $a$, from bottom to top corresponds to the leading (red), next-to-next-to-leading (blue) and next-to-leading order (green) in $F_{W}$ (\eq{weyl}), respectively.  For those converging to $1$ at large $a$, from top to bottom corresponds to the ratio of the numeric force to the large-distance expression (black) \eq{largedistance} and to the next-to-leading order asymptotic expression (grey) --id est, including the two first eigenvalues.  All others shapes (rectangle, triangle, circular and also the non-integrable shapes discussed in next Section) behave qualitatively the same.}
\label{fnumsobrefweyl}
\end{figure}

We have computed numerically the Casimir force $F(a)$ for the three regular pistons and compared it to the predicted behavior for short piston separation, $F_{W}(a)$, and long separation, $F_{\infty} (a)$. The criterion to define whether a separation is long or short depends on the value of $a$ compared to the characteristic lengths of each shape. Since the result for all shapes are qualitatively the same, we have only plotted the results of this comparison for the square with area $A = 1$ in Fig. \ref{fnumsobrefweyl}. (Observe that setting the area to unity is equivalent to use units of $\sqrt{area}$ for the distance $a$ and $1/area$ for the resulting Casimir force.  All figures and results in this paper are to be understood in these units.) As it can be seen in the short-distance region the convergence improves as we include each term of the Weyl expression \eq{weyl}. On the other hand, in the large-distance region, we verify the convergence to the expression in \eq{largedistance} which is a function of the smallest eigenvalue. Moreover, in this region we have also plotted the next-to-leading order large distance asymptotic expression (including the first two eigenvalues) to show explicitly that it is possible --and simple-- to obtain, by choosing different approximations in each region, an analytic approximation of the Casimir force which agrees within
a given accuracy with the numerical result. Indeed, adding a few additional terms in the large distance expansion is possible to describe the numerical results with analytic approximations within a 1\% for any value of $a$.

\begin{figure}[ht]
\psfrag{x}{ $a$}
\psfrag{y}{$F_{num}(a)$}
\begin{minipage}[b]{0.45\textwidth}
\centering
\includegraphics[width=\textwidth]{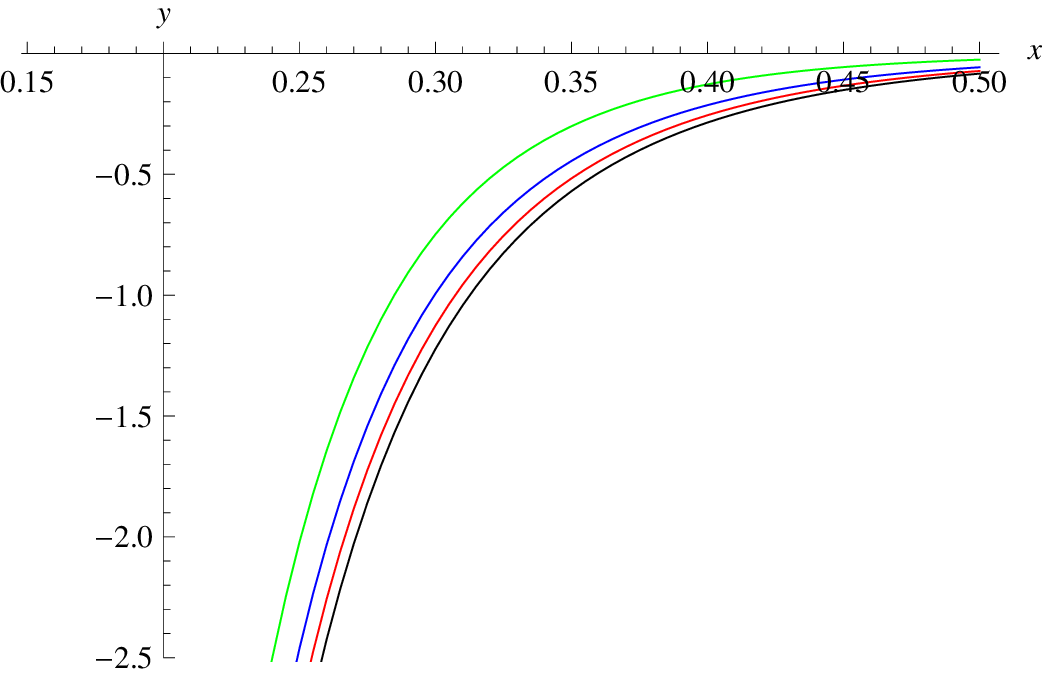}
\end{minipage}
\hspace{0.5cm}
\begin{minipage}[b]{0.45\textwidth}
\centering
\begin{tabular}{|l|c|c|c|}
\mbox{shape}&\mbox{P}&\mbox{$\chi$}&\mbox{$\lambda_1$}\\
\hline
\mbox{circle}&3.54&0.16&4.26\\
\mbox{square}&4&0.25&4.44\\
\mbox{triangle}&4.56&0.33&4.77\\
\mbox{rectangle 4:1}&5&0.25&6.47
\end{tabular}
\vskip 1cm

~

\end{minipage}
\caption{(Color online) Casimir force between integrable-shaped pistons with area equal to unity.  The ordering for small and large $a$ may be understood from the corresponding asymptotic expressions in \eq{weyl} and \eq{largedistance}, respectively, as a function of the ordering in the perimeter and first eigenvalue for each geometry. From top to bottom the lines correspond to a $4:1$ rectangle (green), an equilateral triangle (blue), a square (red) and a circle (black), respectively.  The table shows the different parameters of each geometry that enter into \eq{weyl} and \eq{largedistance}.}
\label{fnum_para_integrables}
\end{figure}

In order to compare the Casimir force for the integrable pistons we have plotted in Figure \ref{fnum_para_integrables} the numeric Casimir force for the different geometries while keeping constant the area $A=1$ for all shapes. Given the short-distance expression $F_{W}(a)$ we find that the differences between the forces at short distances must come from the different perimeters of the pistons. Whereas in the large-distance region the difference must come from the first eigenvalue $\lambda_1$ (see \eq{largedistance}). In table of Figure \ref{fnum_para_integrables} we give the perimeter and first eigenvalue for the geometries under study, as well as the geometrical $\chi$ factor. As it can be seen, there is no cross-over of the lines in Figure \ref{fnum_para_integrables} from large to short distances.  However this is not a general property, since it exists a {\it spectral theory} theorem that states that if region $R$ contains region $R'$ then the lowest eigenvalue of the Laplacian operator restricted to Dirichlet boundary conditions on $R$ is smaller than the lowest eigenvalue of $R'$ \cite{teoremita}.  Therefore, it is possible to imagine pistons $A$ and $B$ where $A$ is contained in $B$, but $A$ has a larger perimeter and, therefore, the ordering in the Casimir forces for short-distances is the contrary to the one for large-distances. We will see an explicit example in Section IV. Another interesting
property is that, among all possible two-dimensional shapes with a given area, the  circle has
the lowest first eigenvalue \cite{teoremita}. Therefore, at large distances the interaction between circular
pistons is stronger than for pistons of any other shape.

We end this Section with some comments on Neumann boundary conditions (TE modes) concerning the pistons studied in this Section. We have computed the Casimir force in each shape for TE modes. The ordering
of the results for small and large distances is the opposite than for Dirichlet boundary conditions (TM modes). At short distances this follows from the expression of $F_{W}$: as already mentioned after equation \eq{weyl}, the term proportional to the perimeter changes sign for Neumann boundary conditions. On the other hand, at large distances the opposite order comes from the relative values of the first eigenvalue of each geometry. The first eigenvalue of the billiards considered here, for Neumann boundary conditions,  is always smaller than the first one for Dirichlet boundary conditions. Therefore the force decreases more slowly as $a \to\ \infty$.

%%%%%%%%%%%%%%%%%%%%%%%%%%%%%%%%%%%%%%%%%%%%%%%%%%%%%%%%%%%%%%%%%%%
\subsection{Correcting part of the force}

To analyze the effects of the classical dynamics on the Casimir force we should investigate the difference

\be
\delta F (a) = F(a) - F_{W}(a) \, ,
\ee
since the Weyl terms $F_{W}$ only depends on geometrical quantities. The difference $\delta F$ is related to the oscillating part of the spectrum and will contain information of classical periodic orbits.  From \eq{weyl}, where the last term proportional to $\chi$ goes as $a^{-2}$, we typically expect that

\bea
\displaystyle\lim_{a\to 0} a^2 \delta F (a) = 0.
\label{condition}
\eea
To see if there is a qualitatively different behavior as $a \to 0$ for different type of dynamics, we study numerically the difference of the force for regular as well as for chaotic (next Section) geometries.

\begin{figure}[h]
\psfrag{x}{$a$}
\psfrag{y}{$\delta F(a)$}
\includegraphics[width=.5\textwidth]{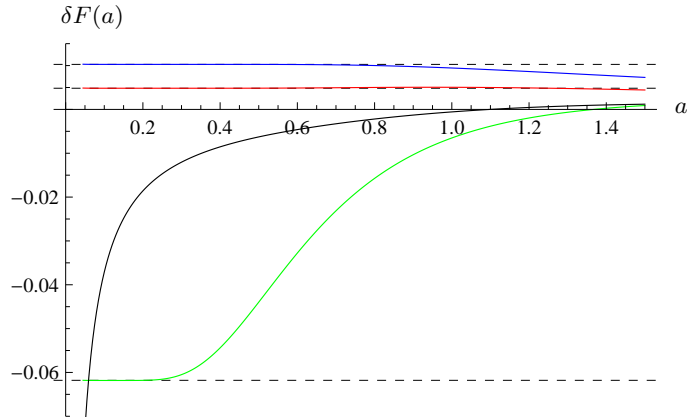}
\caption{(Color online) Correcting part of the force ($\delta F=F-F_{W}$) for integrable geometries. Solid lines from top to bottom at $a=1$ correspond to triangle (blue), square (red), circle (black) and rectangle 4:1 (green), respectively.  The dashed lines are the asymptotes to which the polygon geometries converge according to the explicit calculations in Appendix \ref{apendice}.}
\label{fnummenosfweyl}
\end{figure}

We have plotted $\delta F$ for the integrable shapes studied in Figure \ref{fnummenosfweyl}. As it can be seen, the geometries consisting only in straight lines and angles (square, rectangle and triangle) reach a constant for short distances. For these geometries, semiclassical expansion can be done in an exact way, and the short distance behavior can be obtained explicitly in terms of periodic orbits (see Appendix \ref{apendice}). The constant can be precisely computed and as it can be seen in the Figure (dashed asymptotes) matches perfectly with the numerical evaluation of $\delta F$.  On the other hand, the circle shows a divergence in $\delta F$ for $a\to 0$.  Based on the examples studied in next sections we would conjecture that this is a general property for billiards with curved boundaries.

It is worth mentioning that, had we forgotten to sum a single eigenvalue in \eq{mara} of those required by the accuracy sought (see discussion after \eq{largedistance}), then the condition \eq{condition} would be violated. Indeed, the contribution of
a given eigenvalue $\lambda_k$ to the force is
\be
F_{k}(a)=\frac{1}{\pi} \sum_{l=1}^{+\infty}
\lambda_{k}^2 K'_1(2l\lambda_{k} a) \, ,
\label{oneeigenv}
\ee
and a simple numerical analysis shows that $F_{k}$ diverges as $1/a^2$ for $a\to 0$. Therefore
the wrongly-computed $\delta F(a)$ would diverge as $1/a^2$ violating \eq{condition}.  We have verified
that this is not the case for all the numerical calculations.

%%%%%%%%%%%%%%%%%%%%%%%%%%%%%%%%%%%%%%%%%%%%%%%%%%%%%%%%%%%%%%%%%%%%%%%%%%%%%%%%%%%%%%%%%%%%%%%%%%%%%%%%%%%
%%%%%%%%%%%%%%%%%%%%%%%%%%%%%%%%%%%%%%%%%%%%%%%%%%%%%%%%%%%%%%%%%%%%%%%%%%%%%%%%%%%%%%%%%%%%%%%%%%%%%%%%%%%
%%%%%%%%%%%%%%%%%%%%%%%%%%%%%%%%%%%%%%%%%%%%%%%%%%%%%%%%%%%%%%%%%%%%%%%%%%%%%%%%%%%%%%%%%%%%%%%%%%%%%%%%%%%
%%%%%%%%%%%%%%%%%%%%%%%%%%%%%%%%%%%%%%%%%%%%%%%%%%%%%%%%%%%%%%%%%%%%%%%%%%%%%%%%%%%%%%%%%%%%%%%%%%%%%%%%%%%
%%%%%%%%%%%%%%%%%%%%%%%%%%%%%%%%%%%%%%%%%%%%%%%%%%%%%%%%%%%%%%%%%%%%%%%%%%%%%%%%%%%%%%%%%%%%%%%%%%%%%%%%%%%
%%%%%%%%%%%%%%%%%%%%%%%%%%%%%%%%%%%%%%%%%%%%%%%%%%%%%%%%%%%%%%%%%%%%%%%%%%%%%%%%%%%%%%%%%%%%%%%%%%%%%%%%%%%

\section{Pistons with chaotic classical dynamics} \label{section-chaotic}

\begin{figure}[h]
\includegraphics[width=.5\textwidth]{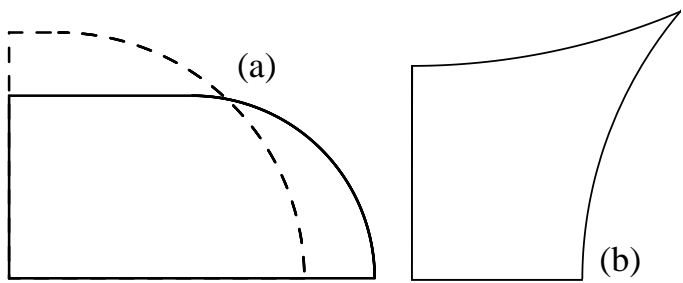}
\caption{Shape of chaotic billiards pistons used in the calculations. (a) Stadiums are a family of figures formed by the union of a rectangle (of sides $r$ and $\ell$) and a quarter of circle (of radius $r$); once the area is fixed to unity they are labeled by the ratio $\ell/r$.  In the Figure we show a stadium with $\ell/r=1$ (solid) and $\ell/r=0.2$ (dashed). (b) Uniformly hyperbolic Sinai-type chaotic billiard, also used in this work with area fixed to unity (see \cite{weird} for details).}
\label{stadium}
\end{figure}

In this Section we compute the Casimir force between pistons with non-integrable shapes. We study a family of stadiums billiards and a Sinai-type billiard. A stadium consist of a rectangle of sides $r$ and $\ell$ with a quarter of circle of radius $r$, as shown in Figure \ref{stadium}. It has been shown that the classical motion of a particle inside a stadium is fully chaotic \cite{fullychaotic}. The second kind of chaotic billiard is also shown in Figure \ref{stadium}. It has been extensively studied in Ref.~\cite{weird}, and its eigenvalues have been computed \cite{eigenvalues}.

The computation of the Casimir force requires the knowledge of the eigenvalues of the Laplace operator with a given boundary condition. The eigenvalues of the chaotic billiards were computed using the scaling method \cite{vergini}. This powerful method was proposed by
Saraceno and Vergini and up to date is one of the most efficients
methods to solve the Helmholtz eigenvalue problem in a hard-walled
billiard (with Dirichlet boundary conditions). Using this method
it is possible to find all eigenvalues and eigenfunctions in a narrow energy range
by solving a generalized eigenvalue problem in terms of quantities over the
boundary of the cavity. The gain being that in a single computational
step a number of accurate eigenvalues are obtained in a constant
proportion to the dimension of the matrices scaling as ${\cal O}(k)$ where
$k = 2\pi/\lambda$ is a referential wave-number.
The success of the method is essentially based on
the fact that eigenstates are quasi-orthogonal on the boundary.
The scaling method was used to numerically compute extremely
high-lying energy levels of $2D$ and  $3D$ billiards \cite{2d,3d}.

\subsection{Casimir force}

We have computed the Casimir force between stadium pistons with Dirichlet boundary conditions for $\ell/r = 1,\ 0.705,\ 0.7,\ 0.205,\ 0.2,\ 0.005$ and $0$, as well as for the Sinai-type billiard, all with area $A=1$ (see Appendix \ref{b} for the details in the samples of eigenvalues).

We have verified the correct convergence of the numerical computation to the expected short and long distance behavior using \eq{weyl} and \eq{largedistance}, respectively, for all billiards. %We show the result for the stadium $\ell/r=1$ in Fig. \ref{fnumsobrefweylSTAD}, all others being qualitatively similar.

%\begin{figure}[h]
%\psfrag{x}{$a$}
%\psfrag{y}{}
%\includegraphics[width=.5\textwidth]{fnumsobrefweylSTAD.eps}
%\caption{(Color online) Idem Figure \ref{fnumsobrefweyl} but for the stadium $\ell/r=1$.  A similar behavior is found for all the non-integrable pistons investigated.}
%\label{fnumsobrefweylSTAD}
%\end{figure}

\begin{figure}[ht]
\psfrag{x}{$a$}
\psfrag{y}{$F_{num}(a)$}
\begin{minipage}[b]{0.45\textwidth}
\centering
\includegraphics[width=\textwidth]{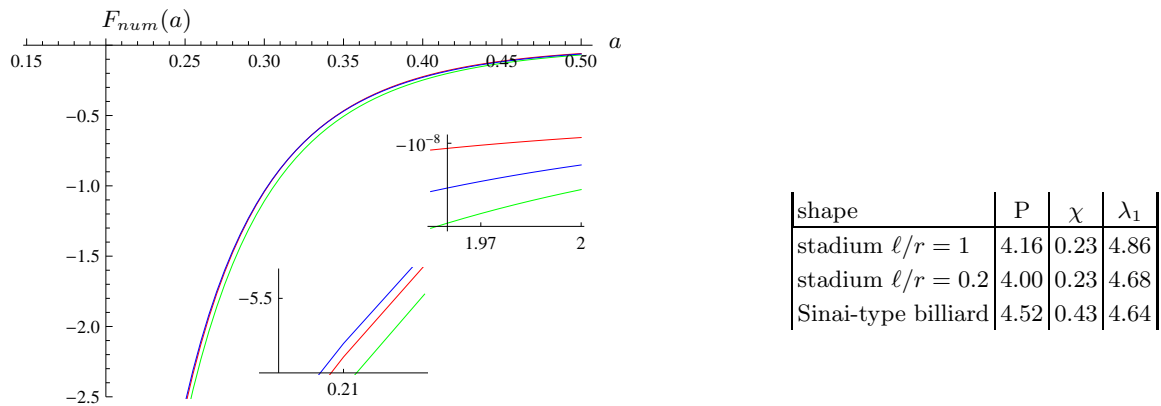}
\end{minipage}
\hspace{0.5cm}
\begin{minipage}[b]{0.45\textwidth}
\centering
\begin{tabular}{|l|c|c|c|}
\mbox{shape}&\mbox{P}&\mbox{$\chi$}&\mbox{$\lambda_1$}\\
\hline
\mbox{stadium }$\ell/r=1$&4.16&0.23&4.86\\
\mbox{stadium }$\ell/r=0.2$&4.00&0.23&4.68\\
\mbox{Sinai-type billiard}&4.52&0.43&4.64
\end{tabular}
\vskip 1cm

~

\end{minipage}
\caption{(Color online) Casimir force for different shaped chaotic pistons. For small $a$ (lower inset) from top to bottom: Sinai-type billiard (blue), stadium $\ell/r=1$ (red) and stadium $\ell/r=0.2$ (green).  For large $a$ (upper inset) from top to bottom: stadium $\ell/r=1$ (red), Sinai-type billiard (blue),  and stadium $\ell/r=0.2$ (green).  Due to the relationship between the perimeters and the first eigenvalue there is cross-over of the forces from the small to the large distance regime. At distances larger than those plotted in this figure there is a crossover between the green and blue curves,
as expected from the fact that the lowest eigenvalue of the $\ell/r=0.2$ stadium is slightly larger than the one
of the Sinai-type billiard.}
\label{fnum_para_NO_integrables}
\end{figure}

We present in Figure \ref{fnum_para_NO_integrables} the numeric Casimir force for three relevant shapes. At this level there are no qualitative differences with the force of integrable geometries. However, in this case, we do see a cross-over of the forces between the Sinai-type piston and the stadiums, since the ordering in the perimeter (see table in Figure \ref{fnum_para_NO_integrables}) is contrary to the order in the first eigenvalue and, therefore, the ordering for the asymptotic small and large distance is reversed (see discussion in previous Section).

\subsection{Correcting part of the force}

\begin{figure}[ht]
\begin{minipage}[b]{0.45\textwidth}
\centering
\psfrag{x}{$a$}
\psfrag{y}{$\delta F(a)$}
\includegraphics[width=\textwidth]{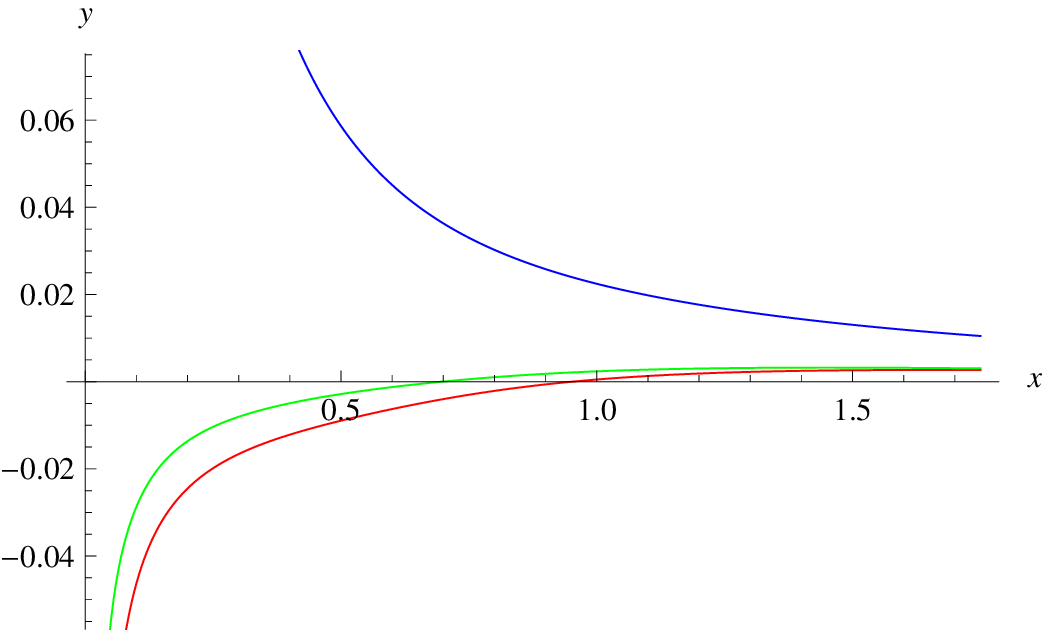}
\newline
(a)
\end{minipage}
\hspace{0.5cm}
\psfrag{x}{$a$}
\psfrag{y}{$a\, \delta F(a)$}
\begin{minipage}[b]{0.45\textwidth}
\centering
\includegraphics[width=\textwidth]{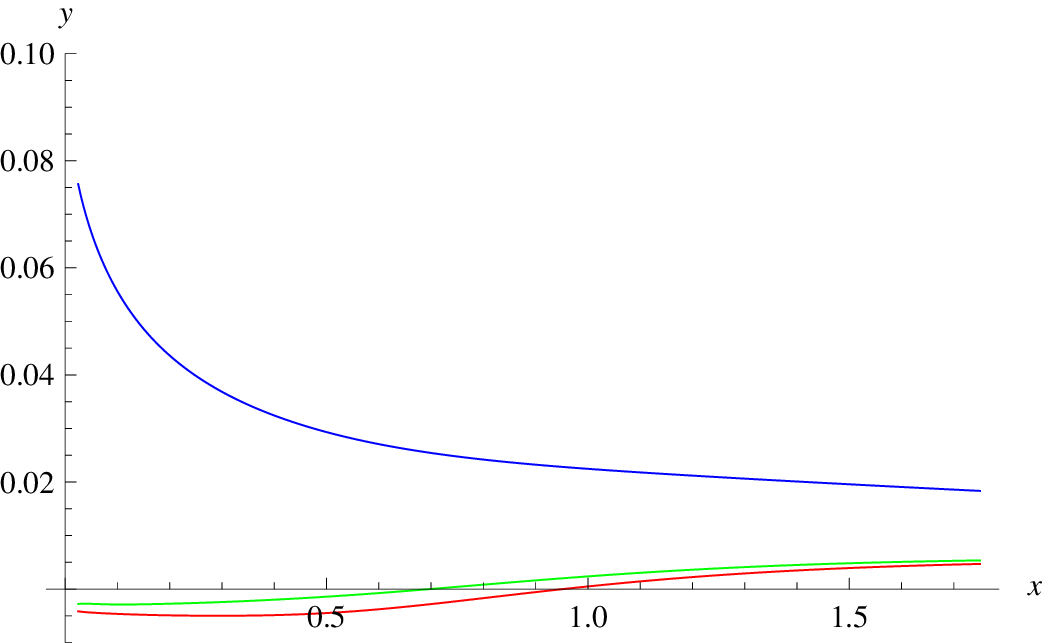}
\newline
(b)
\end{minipage}
\caption{(Color online) (a) Correcting part of the Casimir force for chaotic geometries, all diverge as $a\to 0$. (b) $a\,\delta F(a)$ for these same geometries.  We find that all converge to a constant. On both figures, from top to bottom the plots correspond to the Sinai-type billiard (blue), the stadium $\ell/r=0.2$ (green) and the stadium $\ell/r=1$ (red), respectively.}
\label{oscillatingNI}
\end{figure}

In this Section we investigate the correcting part of the Casimir force that comes from the oscillating part of the spectrum for the non-integrable geometries under study. In all cases we found that in the limit of short distances $\displaystyle\lim_{a\to 0} \delta F (a) = \pm \infty$.  This is contrary to the the straight-line-contours integrable pistons (rectangles and equilateral triangle), but similar to the curved-line-contour (circle). Moreover, as in the circle, for these chaotic pistons with curved contours $\delta F \sim 1/a$ as $a\to0$, because it is observed that $\displaystyle\lim_{a \to 0} a \delta F(a) =$ constant. These results are plotted in Figure \ref{oscillatingNI}. In virtue of the discussion in the previous Section, we verified that $a^2 \delta F(a)\to 0$ as $a\to 0$, and therefore we have not omitted any relevant  eigenvalue.

\section{Chaotic transition study}

In order to look for chaotic distinctive signals we have studied $F(a)$ and $\delta F(a)$ in a transition between chaotic to regular geometries. These correspond to the maximal-chaotic $\ell/r=1$ stadium to the non-chaotic $\ell/r=0$ stadium, which corresponds to the quarter of circle. In this transition, for reasons that will became clear below, we have stepped in the $\ell/r=1,\ 0.705,\ 0.7,\ 0.205,\ 0.2,\ 0.005$ and $\ell/r=0$ (quarter of circle) stadiums.

\begin{figure}[ht]
\begin{minipage}[b]{0.45\textwidth}
\centering
\psfrag{x}{$a$}
\psfrag{y}{$F_{num}(a)$}
\includegraphics[width=\textwidth]{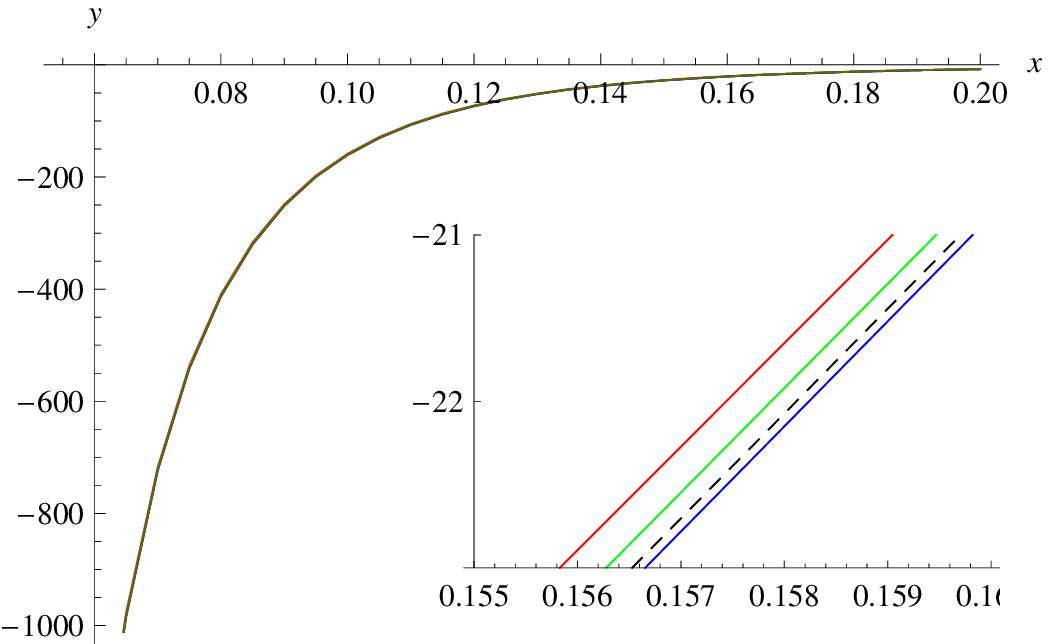}
\newline
(a)
\end{minipage}
\hspace{0.5cm}
\begin{minipage}[b]{0.45\textwidth}
\centering
\psfrag{x}{$\ell/r$}
\psfrag{y}{$perimeter$}
\includegraphics[width=\textwidth]{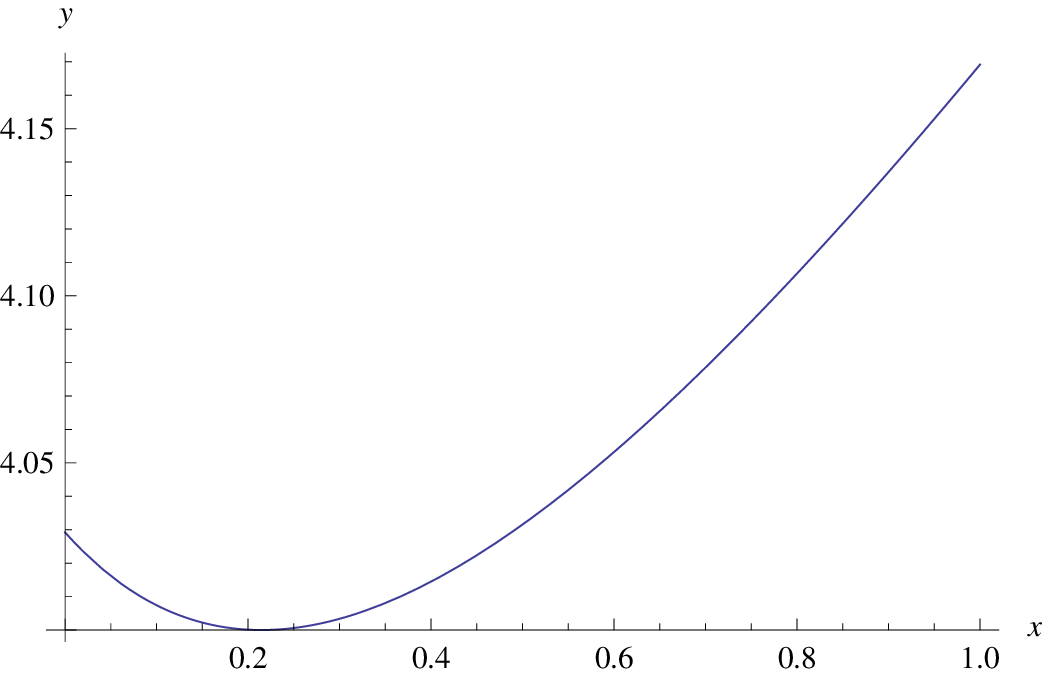}
\newline
(b)
\end{minipage}
\caption{(a) (Color online) Numeric Casimir force for several stadiums.  The lines correspond from top to bottom to stadiums with $\ell/r=1$ (red), $0.7$ (green), $0$ (black dashed) and $0.2$ (blue), respectively.  (b) Perimeter of a stadium of area equal to $1$ as a function of $\ell/r$.  Observe the correlation between both figures, since the perimeter rules the force ordering for the equal-area stadiums.}
\label{peri}
\end{figure}

In a first stage we have computed numerically $F (a)$ and, as expected, we have found no differences in these geometries that could come from a chaotic versus non-chaotic behavior. This would have been the case if the $\ell/r=0$ stadium (quarter of circle) would have had a distinctive behavior in relation to the others $\ell/r\neq0$ stadiums.  We have plotted $F_{num}(a)$ for some relevant stadiums in Figure \ref{peri}.  For small $a$, we have that since all stadiums have the same area and geometrical $\chi$-factor, $F_{num}(a)$ is ruled by the perimeter term in $F_{W}(a)$ (\eq{weyl}). It is interesting to notice that, for fixed area, the perimeter as a function of $\ell/r$ is not a monotone function but, instead, it has a minimum at $\ell/r=1-\pi/4\approx0.21$.  This ordering in the perimeter of the stadiums is also observed in $\delta F(a)$ for large $a$.

\begin{figure}[ht]
%\begin{minipage}[b]{0.45\textwidth}
%\centering
%\includegraphics[width=\textwidth]{fnummenosfweylparatransition.eps}
%\newline
%(a)
%\end{minipage}
%\hspace{0.5cm}
%\begin{minipage}[b]{0.45\textwidth}
\centering
\psfrag{x}{$a$}
\psfrag{y}{$a\, \delta F(a)$}
\includegraphics[width=.5\textwidth]{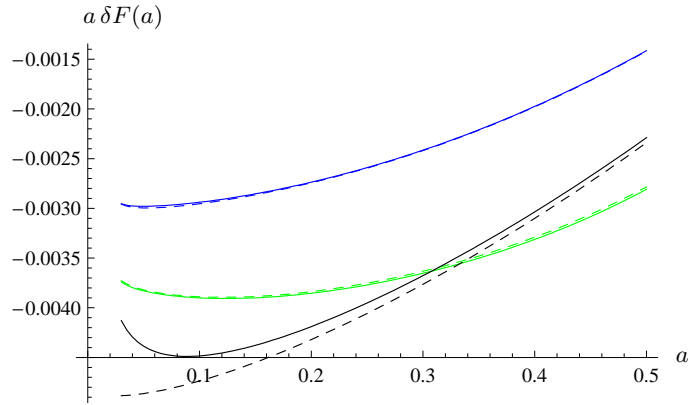}
%\newline
%(b)
%\end{minipage}
\caption{(Color online) $a\,\delta F(a)$ for different values of $\ell/r$ and a tiny variation on it.  From top to bottom: $0.205$ (blue), $0.2$ (blue dashed), $\ell/r=0.705$ (green), $0.7$ (green dashed), $0.005$ (black) and $0$ (black dashed). The plot shows the sensitivity of $\delta F(a)$ in $\ell/r$ at $\ell/r\approx0$, where the transition from non-chaotic to chaotic occurs.}
\label{oscillating2}
\end{figure}

On the other hand, the study of $\delta F(a)$ has shown a qualitative separation between the chaotic stadiums and the non-chaotic $\ell/r=0$ quarter of circle.  Since the expected differences should show up for small $a$ and $\delta F(a)$ diverges as $a\to0$ we found suitable to study this qualitative behavior in the product $a\,\delta F(a)$ where we avoid the divergence. We have found that a small variation of $\ell/r$ produces a small variation of $a\,\delta F(a)$ for $\ell/r\not \approx 0$. On the other hand, for $\ell/r\approx 0$ we have found that $a\, \delta F(a)$ is very sensitive to $\ell/r$.  To show this behavior we have plotted in Figure \ref{oscillating2} the product $a\,\delta F (a)$ for three characteristic values of $\ell/r$ and a tiny variation on each of them.  As it can be seen in the figure, the plots for $\ell/r\not \approx 0$ ($\ell/r=0.2,\ 0.205,\ 0.7$ and $0.705$) show practically no difference, whereas the plots for $\ell/r=0$ and $\ell/r=0.005$ show a large variation.  In other words, the function $U(\ell/r)=\displaystyle\lim_{a\to0} a\,\delta F(a)$ has a smooth behavior at $\ell/r\neq 0$, and a sudden change at $\ell/r=0$.

%%%%%%%%%%%%%%%%   POR ACA VOY   %%%%%%%%%%%%%%%%%%%5

\section{Final remarks}

In this work we have studied numerically the Casimir force between two identical pistons of several shapes inside a cylindrical cavity.  We summarize our main results and conclusions.

We have presented explicit numerical calculations for the force
between pistons of different shapes. The numerical calculations were verified using the large and small distance behaviors, which are known analytically. Moreover, we have shown that it is easy to obtain accurate analytic expressions for the force in the whole range of distances.

For the simplest geometries (rectangles and triangles), the eigenvalues are known analytically, and the evaluation of the force is relatively simple. For the case of the circle, the eigenvalues are known implicitly, and we used two different
methods to compute the sum over eigenvalues: Cauchy's theorem and a numerical evaluation
of the eigenvalues. Finally, for the more complex geometries, the only alternative is to compute the
eigenvalues numerically and then use the exact formula for the force. The evaluation of the force at short distances involves a
very large number of eigenvalues, and therefore a precise and efficient method to compute thousands of eigenvalues is necessary.
We have used a particular method developed in the context of quantum billiards. At this point we would like to
stress that the combination of this or alternative methods to compute eigenvalues \cite{amore} with Cauchy's theorem could be particularly useful to compute the Casimir energy for other geometries (see for instance \cite{pmm}).

One of the original motivations of this work was to look for signals of quantum chaos in the
Casimir force for the piston geometry. For this reason, we evaluated the part of the force originated by the oscillating part of the spectrum. On the one hand,
we have seen a set-apart behavior depending on whether the boundary of the piston  contains curved-lines or not.
In the first case $\delta F$ diverges as $a\to 0$, while in the second case converges.
A more detailed analysis of the transition between chaotic and regular geometries shows a different behavior for the non-chaotic geometry in the part of the force coming from the oscillatory part of the spectrum, more explicitly in the finite limit $U=\displaystyle\lim_{a\to0} a\,\delta F(a)$. We analyzed a one parameter ($\ell/r$) family of stadiums,
that starts with an integrable geometry (a quarter of circle, $\ell/r=0$).  The finite limit is a function of
this parameter, $U=U(\ell/r)$ that has a smooth behavior for $\ell/r\neq 0$, and a jolt at $\ell/r=0$. 

In this paper we analyzed the eventual existence of footprints of the statistical properties of
the spectrum in the Casimir force. There could be signals of quantum chaos
in the fluctuations of the Casimir force. We hope to address this interesting problem in a forthcoming publication.

This work has many numerical results which have given us clues on the behavior of the Casimir Force for different shapes of the pistons. It would be very interesting to explore and verify analytically these results.

%%%%%%%%%%%%%%%%%%%%%%%%%%%%%%%%%%%%%%%%%%%%%%%%%%%%%%%%%%%%%%%%%%%%
%%%%%%%%%%%%%%%%%%%%%%%%%%%%%%%%%%%%%%%%%%%%%%%%%%%%%%%%%%%%%%%%%%%%
%%%%%%%%%%%%%%%%%%%%%%%%%%%%%%%%%%%%%%%%%%%%%%%%%%%%%%%%%%%%%%%%%%%%
%%%%%%%%%%%%%%%%%%%%%%%%%%%%%%%%%%%%%%%%%%%%%%%%%%%%%%%%%%%%%%%%%%%%
%%%%%%%%%%%%%%%%%%%%%%%%%%%%%%%%%%%%%%%%%%%%%%%%%%%%%%%%%%%%%%%%%%%%

\appendix
\section{Eigenvalues samples}
\label{b}

For the different shaped pistons we have used different set of eigenvalues.  The following table summarizes the required information to reproduce our results.

\begin{center}
\begin{tabular}{|c|l|l|}
\hline
{\bf shape} & {\bf total eigenvalues} & {\bf maximum eigenvalue}\\
\hline
square & $30000$ & $615.95$ \\
\hline
rectangle 4:1 & $30000$ & $616.45$\\
\hline
triangle & $61075$ & $963.35$\\
\hline
circle & $19984$ & $549.38$\\
\hline
$1/4$--circle & $16111$ & $451.95$\\
\hline
stadium $\ell/r=1$ & $17247$ & $467.59$\\
\hline
stadium $\ell/r=0.705$ & $17254$ & $467.7$\\
\hline
stadium $\ell/r=0.7$ & $21439$ & $521.08$\\
\hline
stadium $\ell/r=0.205$ & $17258$ & $467.67$\\
\hline
stadium $\ell/r=0.2$ & $18000$ & $477.58$\\
\hline
stadium $\ell/r=0.005$ & $17257$ & $467.63$\\
\hline
Sinai-type billiard & $62076$ & $885.47$\\
\hline
\end{tabular}
\end{center}

%%%%%%%%%%%%%%%%
\section{Correcting part of the force for rectangles and triangles}
\label{apendice}

%\appendix
%\section{Oscillating part of the force for non-curved geometries}

As expression (\ref{mara}) for the Casimir force, the Casimir energy can be rewritten in an
integral form

\be
E^{TM}(a) + E^{TE}(a) = - \frac{1}{2\pi} \sum_{l=1}^{\infty} \frac{1}{l}
\int {\text d} \epsilon \sqrt{\epsilon} K_1 ( 2 l a \sqrt{\epsilon} )
[\rho_{\text D} (\epsilon) + \rho_{\text N} (\epsilon)] \ ,
\label{integral_casimir}
\ee
where

\be
\rho_{\text{D,N}} (\epsilon) = \sum_k \delta (\epsilon - \lambda_{k
\text{D,N}}^2 ) = \bar{\rho}_{\text{D,N}} (\epsilon) +
\widetilde{\rho}_{\text{D,N}} (\epsilon)
\ee
is the density of states with Dirichlet (Neumann) boundary conditions, and
where we have already split the density into a Weyl smooth part plus an
oscillating part. For both rectangular and equilateral triangular
billiards, there are explicit exact formulas for this semiclassical
expansion using Poisson summation formula from explicit expressions
(\ref{eigen_rect}) and (\ref{eigen_tri}) for the eigenvalues (see
\cite{BraBha1997} for details). The smooth part has the usual form for 2D
billiards

\be
\bar{\rho}_{\text{D,N}} (\epsilon) = \frac{A}{4 \pi} \mp \frac{P}{8 \pi
\sqrt{\epsilon}} + \chi_{\text{D,N}} \delta ( \epsilon )
\ee
where $A = L_x L_y, P = 2 (L_x + L_y), \chi_{\text{D}} = 1/4,
\chi_{\text{N}} = -3/4$ for rectangular billiard of sides $L_x$ and $L_y$,
and $A = \sqrt{3}L^2/4, P = 3 L, \chi_{\text{D}} = 1/3, \chi_{\text{N}} =
-2/3$ for equilateral triangular billiard of side $L$. The contribution of
this smooth part of the Weyl series to the Casimir energy can be evaluated
analytically, giving an exact result that has already been discussed, see
\eq{weyl}.

The oscillating parts have the following exact expressions in terms of
periodic orbits

\bea
\mbox{rectangle} \rightarrow \widetilde{\rho}_{\text{D,N}} (\epsilon) &=&
\frac{A}{4 \pi} \sum_{\bf M} J_0 (L_{\bf M} \sqrt{\epsilon}) \mp
\frac{1}{2 \pi \sqrt{\epsilon}} \sum_{m = 1}^{\infty} [ L_x \cos ( 2 L_x m
\sqrt{\epsilon} ) +  L_y \cos ( 2 L_y m \sqrt{\epsilon} ) ]
\label{rho_osc_rect} \\
\mbox{triangle} \rightarrow \widetilde{\rho}_{\text{D,N}} (\epsilon) &=&
\frac{A}{4 \pi} \sum_{\bf M} J_0 (L_{\bf M} \sqrt{\epsilon} ) \mp \frac{3
L}{4 \pi \sqrt{\epsilon}} \sum_{m=1}^{\infty} \cos ( ( 3 L/2) m
\sqrt{\epsilon} ) \label{rho_osc_triang}
\eea
with double index ${\bf M} = (M_1, M_2)$ running over integer numbers, excluding from the sum the term $(M_1,M_2) = (0,0)$, which indeed had
given the area term in Weyl series. The length of periodic orbits are $
L_{\bf M} = 2 \sqrt{(M_1 L_x)^2 + (M_2 L_y)^2} $ for the rectangle, and
$L_{\bf M} = \sqrt{3 (M_1^2 + M_2^2 + M_1 M_2)} L$ for the equilateral
triangle.

We see that for both shapes, and both boundary conditions, the structure
of the formulae are very similar. We have first to evaluate the integral
of terms with $J_0$ function, that can be solved analytically

\be
\int_0^{\infty} {\text d} \epsilon \sqrt{\epsilon} K_1 ( 2 l a
\sqrt{\epsilon} ) J_0 (L_{\bf M} \sqrt{\epsilon}) = \frac{l}{2 a^3 (l^2 +
(L_{\bf M} /2 a)^2 )^2}
\ee
Using this into (\ref{integral_casimir}) the resulting $l$ sum can also be
performed

\be
\sum_{l=1}^{\infty} \frac{1}{ (l^2 + \alpha^2 )^2} = \frac{2 \pi^2
\alpha^2 + 2 \pi \alpha \sinh (\pi \alpha) \cosh (\pi \alpha) - \sinh^2
(\pi \alpha) }{ 8 \alpha^4 \sinh^2 (\pi \alpha)} = \frac{\pi}{4 \alpha^3}
- \frac{1}{2 \alpha^4} + {\cal O} (\ue^{- 2 \pi \alpha} )
\ee
with $\alpha = L_{\bf M} /2 a$. We used the approximation $\alpha \gg 1$,
corresponding to the limit of small plate separation $a$ (compared to the
typical plate's size). The final contribution of the $J_0$ terms to the
Casimir energy is

\be
-\frac{A}{2 \pi} \left( \frac{1}{4}  \sum_{\bf M} \frac{1}{ L_{\bf
M}^3 } - \frac{a}{\pi} \sum_{\bf  M} \frac{1}{ L_{\bf M}^4 }
\right)
\ee
which is the same for TE and TM modes, and both shapes. As a function of
plate separation $a$, there are a constant term plus a linear term on $a$.

For the terms in $\widetilde{\rho}$ with cosines, we have integrals of type

\be
\int_0^{\infty} {\text d} \epsilon K_1 ( 2 l a \sqrt{\epsilon} ) \cos (R m
\sqrt{\epsilon}) = \frac{\pi l}{4 a^2 (l^2 + (R m /2 a)^2 )^{3/2} }
\ee
with $R = 2 L_x$ or $2 L_y$ for the two terms in the rectangular billiard,
and $R = 3 L/2 $ for the equilateral triangle. Now the sum over $l$ can
not be done analytically, but we give an approximation

\be
\sum_{l=1}^{\infty} \frac{1}{ (l^2 + \alpha^2 )^{3/2}} =
\frac{1}{\alpha^2} - \frac{1}{2 \alpha^3} + {\cal O} (\ue^{- \alpha} )
\ee
valid for $\alpha = R m / 2 a \gg 1$. The sum over index $m$ can be done
in terms of Riemann Zeta function. Finally, the contribution of the cosine
terms to the Casimir energy of TM modes (given by the Dirichlet density of
states) is

\bea
\mbox{rectangle} &\rightarrow& \frac{\zeta (2)}{16 \pi} \left(
\frac{1}{L_x} + \frac{1}{L_y} \right) - \frac{\zeta (3)}{32 \pi} \left(
\frac{1}{L_x^2} + \frac{1}{L_y^2} \right) a  \\
\mbox{triangle} &\rightarrow& \frac{\zeta (2)}{6 \pi L} - \frac{\zeta
(3)}{9 \pi L^2} a
\eea
For TE modes the contribution has exactly the opposite sign. We see again
a constant plus a linear term on plate separation $a$. We remark here that
on semiclassical approximation ($\epsilon \rightarrow \infty$), usually
terms coming from cosine in \eqs{rho_osc_rect}{rho_osc_triang} are ignored compared to the main terms from periodic
orbit, and even the $J_0$ function is approximated by its asymptotic
expression for big arguments. But due to the integration from low
energies, we see that both $J_0$ and cosine terms give contributions of
the same order on $a$. This is the reason why a general semiclassical
approximation, only valid for high energies, will not be enough to compute
the contributions to Casimir energy.

Collecting all terms we have the correcting part of the Casimir energy for the rectangle

\bea
\delta E_{\text{D,N}} (a) &=& - \frac{L_x L_y}{8 \pi} \sum_{\bf
M} \frac{1}{ L_{\bf M}^3 } \pm \frac{\pi}{96} \left( \frac{1}{L_x}
+ \frac{1}{L_y} \right) + a \frac{L_x L_y}{2 \pi^2} \sum_{\bf M}
\frac{1}{ L_{\bf M}^4 } \mp a \frac{\zeta (3) }{32 \pi} \left(
\frac{1}{L_x^2} + \frac{1}{L_y^2} \right)
\eea
and for the equilateral triangle

\bea
\delta E_{\text{D,N}} (a) &=& - \frac{\sqrt{3} L^2 }{32 \pi}
\sum_{\bf M} \frac{1}{ L_{\bf M}^3 } \pm \frac{\pi}{36 L} + a
\frac{\sqrt{3} L^2}{8 \pi^2} \sum_{\bf M} \frac{1}{ L_{\bf M}^4 }
\mp  a \frac{\zeta (3) }{9 \pi L^2} \ .
\eea

Deriving with respect to $a$ we finally obtain, up to exponentially small terms, the correcting part of the Casimir force

\bea
\mbox{rectangle} &\rightarrow& \delta F_{\text{D,N}} = - \frac{L_x L_y}{2 \pi^2} \sum_{\bf M}
\frac{1}{ L_{\bf M}^4 } \pm \frac{\zeta (3) }{32 \pi} \left(
\frac{1}{L_x^2} + \frac{1}{L_y^2} \right)  \\
\mbox{triangle} &\rightarrow& \delta F_{\text{D,N}} = - \frac{\sqrt{3} L^2}{8 \pi^2} \sum_{\bf  M} \frac{1}{ L_{\bf M}^4 } \pm \frac{\zeta (3) }{9 \pi L^2} \ ,
\eea
which are constants.

\section*{Acknowledgments}
The work of
E.A., F.D.M.~and D.A.W.~was supported by UBA, CONICET and ANPCyT.  The work of A.G.M.~was supported by CNEA and CONICET.

%%%%%%%%%%%%%%%%%%%%%%%%%%%%%%%%%%%%%%%%%%%%%%%%%%%%%%%%%%%%%%%%%%%%
%%%%%%%%%%%%%%%%%%%%%%%%%%%%%%%%%%%%%%%%%%%%%%%%%%%%%%%%%%%%%%%%%%%%
%%%%%%%%%%%%%%%%%%%%%%%%%%%%%%%%%%%%%%%%%%%%%%%%%%%%%%%%%%%%%%%%%%%%
%%%%%%%%%%%%%%%%%%%%%%%%%%%%%%%%%%%%%%%%%%%%%%%%%%%%%%%%%%%%%%%%%%%%
%%%%%%%%%%%%%%%%%%%%%%%%%%%%%%%%%%%%%%%%%%%%%%%%%%%%%%%%%%%%%%%%%%%%
%%%%%%%%%%%%%%%%%%%%%%%%%%%%%%%%%%%%%%%%%%%%%%%%%%%%%%%%%%%%%%%%%%%%

\end{document}